\documentclass[conference]{IEEEtran}
\usepackage{cite}
\usepackage{amsmath,amssymb,amsfonts}
\usepackage{algorithmic}
\usepackage{graphicx}
\usepackage{textcomp}

 \usepackage{booktabs} 
\usepackage{balance}
\begin{document}

\title{Joint Source-Channel Coding System for 6G Communication: Design, Prototype and Future Directions}
\author{\IEEEauthorblockN{\dag Xinchao Zhong,  \dag Sean Longyu Ma,  Hong-fu Chou, Arsham Mostaani, \\ Thang X. Vu, Symeon Chatzinotas}
\IEEEauthorblockA{\textit{Interdisciplinary Centre for Security, Reliability and Trust (SnT), University of Luxembourg, Luxembourg}\\ 
\textit{\dag School of Computer Science, The University of Auckland, New Zealand}\\ 
Corresponding author: Hong-fu Chou Email: hungpu.chou@uni.lu}

}




\maketitle
\begin{abstract}
The emergence of the AI era signifies a shift in the future landscape of global communication networks, wherein robots are expected to play a more prominent role compared to humans. The establishment of a novel paradigm for the development of next-generation 6G communication is of utmost importance for semantics task-oriented empowered communications. This paper begins by examining the historical development of advanced communications, focusing specifically on the incorporation of semantics and task-oriented features. The goal of semantic communication is to surpass optimal Shannon's criterion regarding a notable problem for future communication which lies in the integration of collaborative efforts between the intelligence of the transmission source and the joint design of source coding and channel coding. The convergence of scholarly investigation and applicable products in the field of semantic communication is facilitated by the utilization of flexible structural hardware design, which is constrained by the computational capabilities of edge devices. This characteristic represents a significant benefit of joint source-channel coding (JSCC), as it enables the generation of source alphabets with diverse lengths and achieves a code rate of unity. Moreover, JSCC exhibits near-capacity performance while maintaining low complexity. Therefore, we leverage not only quasi-cyclic (QC) characteristics to propose a QC-LDPC code-based JSCC scheme but also Unequal Error Protection (UEP) to ensure the recovery of semantic importance. In this study, the feasibility for using a semantic encoder/decoder that is aware of UEP can be explored based on the existing JSCC system. This approach is aimed at protecting the significance of semantic task-oriented information. Additionally, the deployment of a JSCC system can be facilitated by employing Low-Density Parity-Check (LDPC) codes on a reconfigurable device. This is achieved by reconstructing the LDPC codes as QC-LDPC codes. The QC-LDPC layered decoding technique, which has been specifically optimized for hardware parallelism and tailored for channel decoding applications, can be suitably adapted to accommodate the JSCC system. The performance of the proposed system is evaluated by conducting BER measurements using both floating-point and 6-bit quantization. This is done to assess the extent of performance deterioration in a fair manner. The fixed-point system is synthesized and subsequently used to a  semantic feature transmission and reception system across a noisy channel, with the aim of presenting a prototype for semantic communications. This study concludes with some insights and potential research avenues for the JSCC prototype in the context of future communication.
\end{abstract}

\maketitle

\section{Introduction}
\label{sec:introduction}
Traditional communication systems often overlook the significance of the meaning behind information. They operate on the assumption that all symbols or bits are of equal importance and are handled as such. The primary objective of these systems is to ensure the accurate retrieval of transmitted sequences at receiving ends, prioritizing conformity in transmission. The design methods in this field have predominantly relied on principles from digital communication. Information theory establishes the maximum limits on the capacity of the system. While channel coding concentrates on developing strategies that can approach these limits with extremely low error probability, source coding(known as data compression) refers to the process of encoding information in a way that reduces the amount of data required with the objective of optimizing the length of the source encoded sequence. However, the latest generation of communication systems is being applied in ways that challenge the conventional design paradigm, particularly in terms of semantic and task-oriented aspects\cite{mostaani2023taskoriented}. 

The semantic aspect pertains to the level of precision and accuracy with which transmitted symbols are able to convey the intended meaning and comprises the transmission of a notion or informational material from a source to a destination without delving into the intricacies. It entails the comparison of the inferred meaning by the recipient with the intended meaning by the sender while considering the content, requirements, and semantics in order to enhance the communicative system towards a state of intelligence. Furthermore, the task-oriented aspect prioritizes task completion and efficiency and examines the potential ramifications associated with the utility of the provided information. The efficacy of a task or performance metric is determined by how efficiently the acquired information aids in its accomplishment. The achievement of a shared aim within task-oriented limitations and requirements is facilitated by the utilization of available resources, including communication bandwidth, computing expense, and power consumption. The evaluation of system performance can be measured in relation to the extent to which a certain objective is achieved, taking into account the allocated resources, instead of considering all the transmission sequences that can typically be conveyed in the aforementioned information theory framework-based approach. Based on the stated task objectives and existing knowledge, it can be demonstrated that semantic source coding \cite{Lu2022SemanticsEmpoweredCA} has the potential to achieve greater reductions in redundancy expense and communication overhead compared to the Separate Source-Channel Coding (SSCC). This is mostly due to its ability to refine the most pertinent and concise information and then condense it. An intriguing aspect worth exploring is the extent of compression achieved in semantic source coding in relation to the original information\cite{tang2023informationtheoretic}. The authors have obtained the theoretical boundaries for lossless and lossy compression based on this semantic source, together with the lower and upper limitations on the rate-distortion function. 

Furthermore, semantic-information feature learning \cite{Xu2022} is the key to addressing the utilization of cognitive techniques employed as a means to direct computational resources towards activities of higher importance. The process of achieving dynamic adaptation in semantic compression involves the utilization of a feature learning module and an attention feature module. These modules enable the source encoder to generate a variable number of symbols and modify the capability of the source encoder and channel encoder through the use of cognitive techniques. Moreover, task-oriented feature learning\cite{Shi2023} addresses that the attainable precision of inference is contingent upon the amount of feature components collected and the extent to which they are distorted by detecting noise and quantization defects. The classification gain is only influenced by the distributions of classes in the feature space and represents the highest possible inference accuracy that can be attained theoretically. Therefore, this learning accuracy is dependent on not only the underlying design of the artificial intelligence models being used but also the classification distributions of the feature space. In order to facilitate the implementation of immediate intelligent services in future communication systems, it is advantageous to distill and communicate merely the information that is pertinent to the task at hand and precise semantics. The semantic and task-oriented approach effectively reduces the overall latency of the system. However, it should be noted that this method deviates from the optimal Shannon's SSCC \cite{shannon}, which is typically applied in the communication of long block-length bit sequences. SSCC employs advanced compressing methods to eliminate all redundant sequences from source-encoded symbols, whereas Joint Source Channel Coding (JSCC) exploits the remaining redundant information that emerges following compressing with the error-correcting capability in order to minimize distortion within a certain limit of codelength. Unequal error protection (UEP) prevents errors from occurring by assigning encoded redundancies according to the significance of the information bits. Only some sets of bits are of equal significance when sending source-encoded information because of the varied degree of vulnerability of the source decoder. Therefore, the authors in \cite{he2023rate} present a remarkable performance of UEP JSCC code system by providing an innovative adjustable code rate for multiple semantic task classes. We summarize our contribution from the following perspectives:
\begin{enumerate}
    \item This paper provides a brief exploration of the latest designs and methodologies in semantic and task-oriented communication, with a specific emphasis on the knowledge pertaining to the prototype of the JSCC scheme. The proposed fixed-point JSCC scheme serves as a practical solution not only for transmitting and receiving semantic features over noisy channels but also for UEP semantic importance applications. This study presents a novel approach for enhancing the semantic encoder/decoder by including the UEP capabilities of the quasi-cyclic Low-Density Parity-Check (QC-LDPC) JSCC decoder. The primary objective is to safeguard the semantic significance of 6G communication.
    \item The primary aim of this study is to explore the potential of surpassing the traditional Shannon criteria, while also presenting the initial iteration of a prototype for a semantic JSCC communication system. The JSCC prototype significantly reduces the data width from 32-bit floating-point to 6-bit fixed-point, making practical FPGA implementation possible. Additionally, we revisit the design of semantic codec learning and task-oriented signal compression in order to explore how the semantic task-oriented techniques can be adapted to the proposed JSCC platform. 
    \item The JSCC system under consideration acts as a prototype to make it easier to modify communication protocols in the future. Through the use of deep learning techniques, this system is specifically created to be adaptable to a broad variety of semantic and task-oriented features. The proposed prototype, when compared to other state-of-the-art, can deliver better BER performance despite the reduction in fix-point implementation and is achieved by applying QC-LDPC codes with a reasonable code rate.
\end{enumerate}

In this paper, we revisit the design methodologies of next-generation 6G communication systems with regard to semantic and task-oriented aspects and JSCC system following the demonstration of the state-of-the-art prototype designs in Section II.  Second, we investigate the applications of a JSCC system based on QC-LDPC codes. By leveraging quasi-cyclic characteristics and optimizing the whole system, QC-LDPC codes are adopted to enable the feasibility of a JSCC system. We deploy the JSCC system in a Field-Programmable Gate Array (FPGA)-based platform so the users can configure this platform after manufacture.  We also demonstrate that semantic feature transmission and reception are potential application scenarios. The JSCC system can maintain a high code rate (0.8) at a low level of $E_{b}/N_{0}$, which tandem coding systems must decrease to no more than 0.5. In general, the flexibility and performance enhancement of the JSCC system is particularly attractive to next-generation 6G communication. In section III, the proposed QC-LDPC code-based JSCC system is detailed. Section IV presents the experiments and the corresponding results followed by not only a demonstration of a semantic feature transmission and reception but also applicable to task-oriented features. The primary goal is to provide a prototype venture for semantic communications to the marketplace for consumers. Lastly, conclusions and future directions are depicted in Section \ref{sec:conclusion}.
\section{Design of Next Generation Communication and Joint Source Channel coding}
In this section, we examine the application of JSCC system in relation to the integration of AI for semantic and task-oriented considerations to explore the opportunities of future communications.
\subsection{Semantic Codec learning}
\begin{figure*}[htbp]
\centerline{\includegraphics[width=\textwidth]{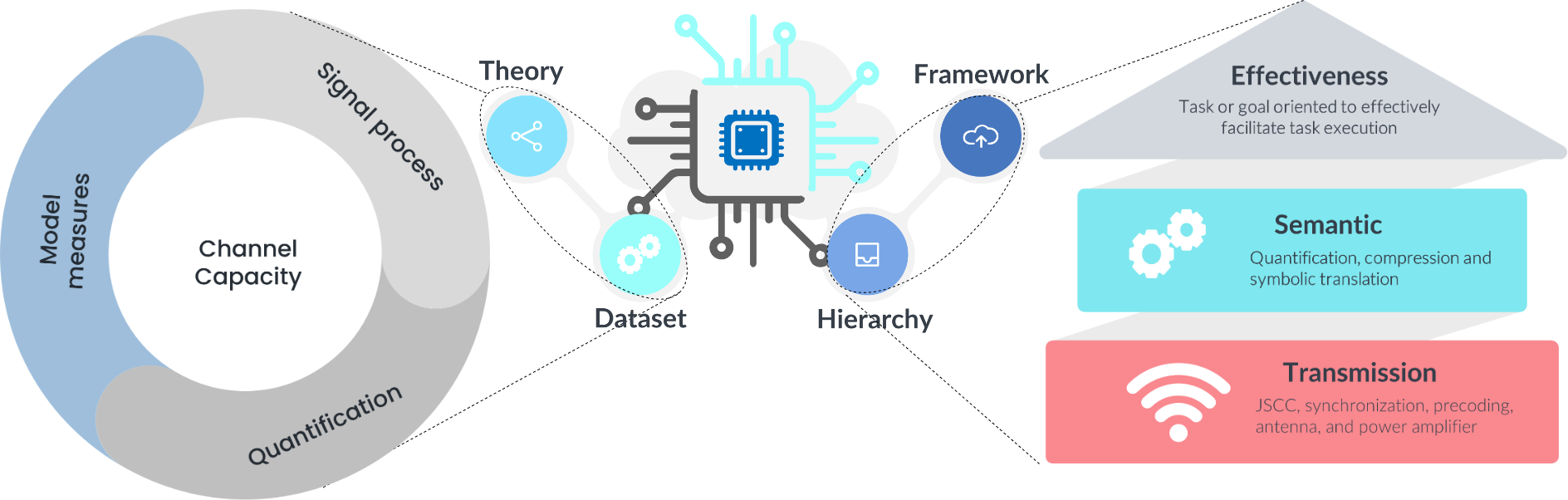}}
\caption{The design overview for semantic communications in \cite{Lu2022SemanticsEmpoweredCA}}
\label{JSCC_sem}
\end{figure*}
The evolution of semantic communication \cite{Lu2022SemanticsEmpoweredCA} is traced back to the early 20\textit{th} century with its continuous growth into the realm of modern communications with regard to beyond 5G and 6G technologies. In light of this, there is a significant need to develop more intelligent and efficient communication protocols that can meet the diverse quality of service (QoS) needs. This must be done while addressing the challenge of limited communication bandwidth and computation. The development of an intelligent communication system is considered essential in both industry and academia. Such a system is not limited to memorizing data flows based on rigorous regulations but also aims to comprehend, analyze, and articulate the fundamental semantics. The ambition of "beyond Shannon" \cite{strinati20216g} surpasses the conventional Shannon paradigm, which focuses on ensuring the accurate receipt of individual sent bits, regardless of the conveyed meaning of these bits. 

In the context of conveying meaning or achieving a goal through communication, the crucial factor lies in the influence exerted by the received sequences on the interpretation of the message meant by the sender or on the attainment of a shared objective. However, despite its growing popularity, research on semantic communication remains fragmented and encompasses a wide range of research interests. The evolution of human dialogues in relation to the semantics of everyday usage for the purpose of semantic communication is still in its infancy. The challenge of developing theoretical semantic models for actual multidimensional information has led to the adoption of JSCC technique in most extant semantic designing strategies. The current implementation of module architectures, similar to the classical Shannon paradigm, presents several issues. As the level of interest in this particular field continues to increase, there is a concerted effort being made to address and surmount the challenges associated with it. Hence, the semantic scheme based on JSCC emerges not only as a highly viable contender for next-generation 6G communication systems but also as a promising transit candidate for the optimal goal of semantic communication. The partnership between JSCC and deep learning in \cite{farsad2018deep} reveals that the deep learning encoder and decoder, as presented, exhibit superior performance in terms of word error rate compared to the conventional technique, particularly when the computational resource allocated for syllable encoding is limited. A limitation of this approach is the utilization of a predetermined bit length for encoding words of varying lengths. In \cite{Xie2021}, a performance comparison of deep and JSCC-based semantic communication\cite{farsad2018deep} to present the potential advantage and the design strategies has concluded the difference between semantic communication and traditional communication as follows:
\begin{enumerate}
\item There exist various domains of processing information. The first phase of SC involves the manipulation of information within the semantic realm, whereas the conventional focuses on compressing information within the realm of entropy.
\item Traditional communication methods prioritize the precise retrieval of information, whereas semantic communication systems are designed to facilitate decision-making or the achievement of specific transmission objectives.
\item Conventional systems primarily focus on designing and optimizing the information transmission modules found in standard transceivers. In contrast, semantic communication systems take a holistic approach by jointly designing the entire information processing framework, spanning from the source information to the ultimate goals of applications.
\end{enumerate}
In comparison\cite{Xie2021}, the complexity analysis of the deep-SC scheme demonstrates superior performance compared to existing SSCC schemes while the JSCC-based semantic communication scheme has lower computing latency than the deep-SC scheme.
The successful incorporation \cite{Xie2021} and the survey \cite{Lu2022SemanticsEmpoweredCA} of deep learning and JSCC techniques inspires the innovative construction of semantic communication. 
The subject matter pertains to the examination of semantic source coding\cite{tang2023informationtheoretic} in connection with the primary information, thereby enabling an inclusive research framework that has the capacity to surpass the limitations of Shannon's conventional information theory. 

In Fig.\ref{JSCC_sem}, we summarize the design overview of the semantic aspect in \cite{Lu2022SemanticsEmpoweredCA} from semantic theory encircled semantic channel capacity and future communication channels and datasets, including text, audio, image, video, unmanned aerial vehicles (UAVs), and the Internet of Things (IoT), are based on semantic theory. In order to calculate the source semantic entropy, a model measurement is required. The semantic quantification determines the semantic compression and signal processing. Furthermore, the semantic framework and hierarchy are illustrated in Fig.\ref{JSCC_sem}. To effectively enable task execution, the effectiveness on the top of the design architecture is based on task or goal-oriented methodologies. Following the basics of theory and dataset, the second level brings the semantic aspect to an evolution of future communication. For the bottom level, this layer via physical transmission integrates with the semantic level and presents JSCC encoder/decoder regarding synchronization, precoding, antennas, and power amplify. 
\subsection{Task-Oriented signal compression}
\begin{figure*}[htbp]
\centerline{\includegraphics[width=0.8\textwidth]{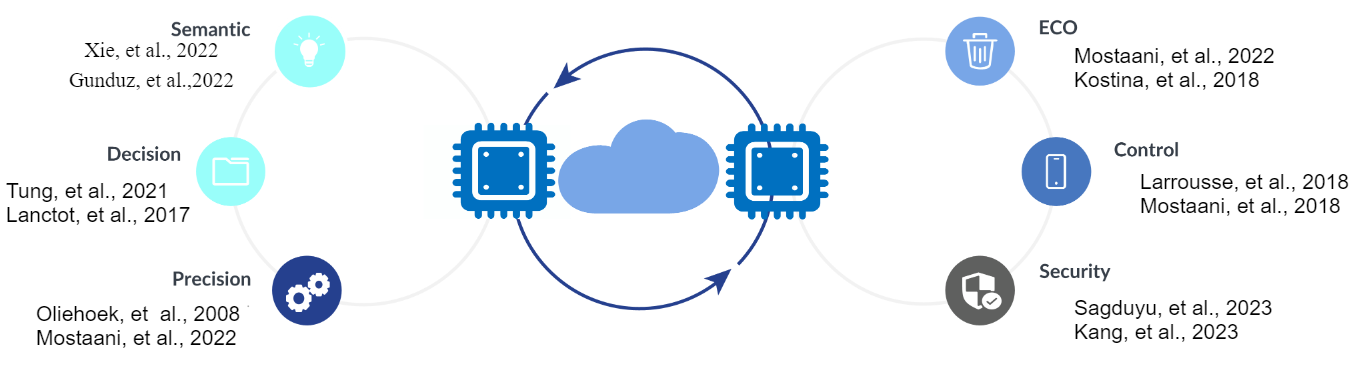}}
\caption{Selected reference of the machine-to-machine communications for task-oriented design}
\label{JSCC_task3}
\end{figure*}
The tremendous growth seen in telecommunication technologies is yet to follow the goal of reliable, fast, and low latency communications or to improve the capacity at which a communication network can serve users. While the idea of reliable communication appears to be a very intuitive and obvious requirement for communication systems, it is arguably a human need.  
Be it communication for a voice call, download of a photo, or streaming a video, we always prefer to receive our desired content at the best perceivable quality. This need, however, is no longer in place when designing machine-to-machine communications. The machine-to-machine communications \cite{mostaani2023taskoriented} occur since this can help the receiver to make more informed decisions 
\cite{mostaani2022task,tung2021effective,mostaani2019Learning} or more precise estimates or computations\cite{shlezinger2021deep} or both \cite{gutierrez2022intent}. Naturally, in this context, there is no need for the reliability of communications to be beyond serving the specific needs of the control, estimation, or computational task at hand. This calls for a fresh examination into the design of communication systems that have been engineered with reliability as one of their ultimate goals \cite{mostaani2021task}. The emerging literature regarding SC \cite{agheli2022semantic} as well as goal/task-oriented communications \cite{mostaani2022task} is attempting to take the first steps towards the above-mentioned goal, i.e., incorporating these semantics\cite{gunduz2022beyond,xie2022task}, together with the goal of message exchange, into the design of communication systems. The ever-increasing growth of machine-to-machine communications is the major motivating factor behind the accelerating research interest in the task-oriented design of communications. As IoT networks and cloud-based applications become more commercialized, autonomous vehicles/UAVs become more mature, and industry 4.0 approaches maturity, a boom in machine-to-machine communications is fueled. To emulate a cyber-physical system composed of several inter-dependant devices or machines, this paper considers the mathematical framework of a generalized decentralized partially observable Markov decision process (Dec-POMDP). There is a significant body of literature behind the theoretical advancements for solving generalized forms of Markov Decision Processes \cite{oliehoek2008optimal}, and their applications in telecommunication and cyber-physical systems \cite{lanctot2017unified,tung2021effective}. Their work departs from the literature on the instance design of the observation function for each agent in the Dec-POMDP. The challenge of jointly developing the observation function and control strategy for each agent in Dec-POMDPs was investigated in \cite{Mostaani2022}. It is important to note that the agents' observations come from a fundamental Markov decision process (MDP). While in classical Dec-POMDP problems \cite{oliehoek2008optimal}, the observation function is considered to be a single fixed function, the framework in \cite{Mostaani2022} offers more flexibility in designing the control policies for a multi-agent system. This approach specifically permits a restricted joint design of the observation and control policy, which is summed up as follows: 
\begin{enumerate}
\item The bit-budgets for the inter-agent communication channels are respected; 
\item The observation functions filter any non-useful observation information for each agent; 
\item The removal of non-useful observation information by the observation functions is carried out such as minimization of any loss on the average return from multi-agents system (MAS's) due to bit-budgeted inter-agent communications. 
\end{enumerate}
The approach in \cite{Mostaani2022} is neither a classic MDP nor a POMDP \cite{monahan1982state} as the action vector is not jointly selected at a single entity: a task-oriented data compression (TODC) problem\cite{kostina2018ratecost,Mostaani2022} can be approximated by identifying the quantization policy in the joint control and quantization problem. A limited bit-budget for the multi-agent communication channels can be achieved with the aforementioned approaches to maximize the expected return by the system. The analytical investigation was presented in \cite{mostaani2018learning,Larrousse2018,Mostaani2022} into how the TODC can be disentangled from the control problem - given the possibility of a centralized training phase. The author's analytical studies confirmed that despite the separation of the TODC and the control problems, they can ensure very little compromise on the average return by the MAS when compared with jointly optimal control and quantization. It is worth noting that the conventional quantization problems regard minimizing the absolute difference between the original signal and its quantized version. However, the difference between task-oriented communication is achieved by considering the usefulness and value of the goal-oriented approach for the task at hand, while conventional communication does not consider it. 
The significance of the result obtained from \cite{Mostaani2022} is multi-fold: 
\begin{enumerate}
\item  Reduces the complexity of the clustering algorithm by transforming it from multi-dimensional observations to the one-dimensional output space of the value functions, \item  The observation points are linearly separable when being clustered according to the generalized data quantization problem 
\item The effectiveness of the data for the task is considered for goal-oriented quantization. 
\item  The value of the observations begins to grow as the ultimate target of the task at hand becomes closer.
\end{enumerate}
Furthermore, the prevalence of deception and Trojan assaults utilizing adversarial machine learning poses a serious threat to machine-to-machine communications and edge servers/devices in \cite{sagduyu2023taskoriented,sagduyu2023vulnerabilities,Kang2023AdversarialAA}. The authors demonstrate adversarial threats and the potential methodology for encouraging more approaches with security and defense of task-oriented communications on 6G networks. In Fig.\ref{JSCC_task3}, the aforementioned design methodologies are summarized as the selected reference for the task-oriented aspect of next-generation 6G communication. Therefore, the previously discussed principles on goal-oriented quantization can be effectively employed in the JSCC scheme to achieve further resource optimization.

\subsection{Joint Source Channel coding and its Prototype}
\begin{table*}
\caption{The significant prototype advances made in the domains of JSCC and DJSCC.}
\centering
 \begin{tabular}{||c c c c c c c ||} 
 \hline
 Paper & \begin{tabular}[c]{@{}c@{}}Implementation\end{tabular} & \begin{tabular}[c]{@{}c@{}}Resource \\ Analysis\end{tabular} & Evaluation & \begin{tabular}[c]{@{}c@{}}Source/Channel\\  coding\\ Quantization\end{tabular} & \begin{tabular}[c]{@{}c@{}}Prototype\\ architecture\end{tabular} & \begin{tabular}[c]{@{}c@{}}Analytical/\\ Data-driven \end{tabular}  \\ [0.5ex]  \hline\hline
 \small{\cite{Ullah2016}} & SDR FPGA &  \small{$\times$} & Experimental BER & BCJR USRP floating & Implicit & Data-driven \\
  \hline
 \small{\cite{Brejza2017}} & FPGA &  \small{$\checkmark$} & \small{Fix-point BER/SER} & Turbo fix-points & Fully parallelism  & Data-driven \\ 
 \hline
  \small{\cite{romero2014analog}} & VLSI Analog &  \small{$\checkmark$} & Experimental SDR & LDPC Analytical & Implicit & Analytical \\ 
 \hline
   \small{\cite{zhao2016low,zhao2017improved}} & LTSpice &  \small{$\checkmark$} & Experimental SDR & N/A & Implicit & Data-driven \\ 
 \hline
   \small{Our work} & FPGA &  \small{$\checkmark$} & fixed-point BER & \small{QC-LDPC 6 bits} & Partial parallelism  & Data-driven \\ 
 \hline
   \small{DJSCC\cite{Liu2022}} & SDR FPGA &  \small{$\times$} & Experimental PSNR & USRP floating & Jpeg CNN & Data-driven \\ 
 \hline
    \small{DJSCC\cite{Yoo2023}} & SDR FPGA &  \small{$\times$} & Experimental PSNR & USRP floating & ViT CNN & Data-driven \\ 
 \hline
\end{tabular}
\label{table: related-works}
\vspace{-3mm}
\end{table*}
In accordance with Shannon's separation theorem, a typical cascade structure that improves the performance of source coding and channel coding independently may maintain the entire system at its optimum\cite{shannon1948mathematical}, such as the majority of contemporary systems for wireless image transmission, which compresses the picture using a source coding method (e.g., JPEG, WebP, BPG) before encoding the bit stream with a source-independent channel code (e.g., LDPC, Polar, Turbo, BCH etc.). Nevertheless, the theory is based on certain premise conditions, such as an unlimited code length, point-to-point transmission system, memory-less stationary source, etc. Since these requirements are seldom satisfied in real-world applications, these tandem coding schemes are often suboptimal, such as autonomous driving and the Internet of Things (IoT) that enforce low latency real-time communication and/or low computation complexity implementation. In addition, if the channel quality goes below a specific level, the channel coding may not offer enough error corrections, and the source coding will inevitably collapse catastrophically. Consequently, jointly optimizing source coding and channel coding for relatively short messages, also known as JSCC, gradually becomes advantageous and garners a great deal of interest. 

JSCC was initially conceptualized more than four decades ago \cite{mceliece2004theory}. This has been explored further since the 1990s\cite{sayood1991use}\cite{hagenauer1995source}\cite{jeanne2005joint}. An iterative joint source-channel decoding algorithm was then proposed in the subsequent works, such as \cite{zribi2012low}, and it was verified that these structures produce a large coding gain over separate coding in finite block-length transmission. 

The article~\cite{fresia2010joint} presented a novel JSCC scheme in which double LDPC codes~\cite{gallager1962low} were applied as the source and channel codes. It was also found that for fixed-size blocks, the source (encoded or unencoded) is redundant, and this redundancy may be exploited on the decoder side. For instance, the channel encoder uses information from the source to lower its frame error rate (FER) while maintaining a very low signal-to-noise ratio (SNR).

Image transmission and reception in a JSCC system were soon proposed as a feasible application. The authors of \cite{bursalioglu2013joint} explored the possibility of transmitting images in a JSCC system through a deep-space communication channel. The authors of \cite{gabay2000real} proposed a joint source-channel coding scheme using BCH codes in a binary symmetric channel (BSC), and it reduced the distortion of satellite images better than a classical tandem source-channel coding scheme based on BCH codes.

In \cite{xu2019joint},\cite{lau2021joint}, the authors optimized the JSCC system or proposed new codes to enhance systemic performance, such as the BER. The research on JSCC systems has become more popular~\cite{4129333, 4537466, 4571120, 7452543, 8016398, 9115063}, but the feasibility of implementing a JSCC system at the circuit level is rarely mentioned. As a sub-class of LDPC codes~\cite{gallager1962low, slepian1973noiseless, 4411018, 5768345, 8293845}, QC-LDPC, where the parity check matrix is composed of permutation matrices (CPMs), can contribute to effective partial-parallel processing, due to the regularity of their parity check matrices \pmb{H}. For telecommunication, there have been comprehensive studies~\cite{7932182, 9179021, 9215273} on improving the complexity and accuracy of the LDPC decoders. They applied appropriate calculations~\cite{6419075, 7388304, hu2001efficient} and various structures of LDPC codes~\cite{6131118, 6481477, 7370816, 8629018}. Other than communications, LDPC code decoders have also been popular in other areas such as storage~\cite{1706484, 7804048, 9015393} or biometric systems~\cite{uludag2004biometric, sutcu2008feature, 9622068}.

The prototype of the JSCC decoder can be remarked on and summarized in \cite{Brejza2017} that trace back to the early 20th century the initiative of JSCC implemented in Variable Length Error Correction (VLEC) code\cite{Buttigieg1994} exhibits a considerable level of intricacy due to the utilization of a vast alphabet for the selection of encoded sources. With the advent of the UEC-URC code\cite{Maunder2013}, Unary Error Correction (UEC) code combined with a Unity Rate Convolutional (URC) code. It provides a nearly optimal performance at a reduced computational cost. This holds true even when dealing with extensive encoded sources, such as those seen in source coding. Although the Log-BCJR algorithm used in UEC-URC decoding is hindered by the presence of sequential information dependencies, which negatively impact processing latency, a solution known as the Fully Parallel Turbo Decoder (FPTD)\cite{Maunder2015} overcomes these limitations. By eliminating the aforementioned dependencies inherent in the traditional Log-BCJR approach, the FPTD attains the performance of the first high throughput near-capacity JSCC decoder and its prototype architecture\cite{Brejza2017}. Furthermore, the Decode-or-Compress-and-Forward (DoCF)\cite{Ullah2016} is proposed for the mechanism that the JSCC decoding step is responsible for processing the demodulated sequences from a relay, followed by an additional stage to retrieve the original message from the source, taking into account the fluctuating circumstances of the channel. The method of the decoder at the point of arrival involves a two-phase decoding procedure that utilizes the standard BCJR algorithm. The experimental verification of the proposed scheme's superiority is conducted by implementing it on Software-Defined Radios (SDRs), with a focus on addressing system-level difficulties in provisioning. The analog circuit prototype\cite{romero2014analog,zhao2016low,zhao2017improved} of JSCC can get a compression technique for lightweight devices and the performance of the system was assessed through the utilization of Spice simulations, in addition to the construction and testing of PCB prototypes. The variant of deep learning-based JSCC (DJSCC)\cite{kurka2021bandwidthagile,tung2022deep,Kurka2022,xie2023robust} is to attain notable levels of reliability despite constraints such as restricted resources and poor SNRs. The reconstruction work involves the utilization of DJSCC, which may be seen as a variant of an auto-encoder. In DJSCC, the SNRs are introduced into the intermediary segment of the encoder, resulting in a distorted version of the original auto-encoder. The DJSCC encoder employs semantic signals or analog waveform sequences instead of digital transmission, unlike traditional wireless transmission. This enables its operation in challenging channel circumstances while still achieving satisfactory restoration. Nevertheless, the efficiency enhancement obtained by current DJSCC techniques is only discernible through simulations in academic studies, which serve as the knowledge sharing by auxiliary transmission for semantic networks\cite{Shi2023}. These simulations typically assume ideal synchronization, precoding, antenna, and power amplifier conditions. Therefore, the study\cite{Liu2022} focuses on an SDR-based DJSCC platform. The capability of this system is evaluated by taking into account two important factors: synchronization error and non-linear distortion. Moreover, drawing inspiration from the impressive resilience demonstrated by Vision Transformers (ViTs)\cite{yoo2022demo} in effectively addressing various challenges associated with picture nuisances, the authors in \cite{Yoo2023} provide a novel approach that utilizes a ViT-based framework for the purpose of SC.
The methodology employed in our study demonstrates an increase in peak signal-to-noise ratio (PSNR) by a satisfactory level when compared to several variations of convolutional neural networks. Finally, we summarize the state-of-the-art prototype of JSCC and DJSCC in Table \ref{table: related-works}. Our work demonstrates the inclusive prototype of an edge semantic device to inspire further advanced hardware design for JSCC-based semantic task-oriented communication.

\section{Proposed Prototype of JSCC system for semantic feature transmission and reception }
\subsection{Overview}

\begin{figure*}[htbp]
\centerline{\includegraphics[width=1\textwidth]{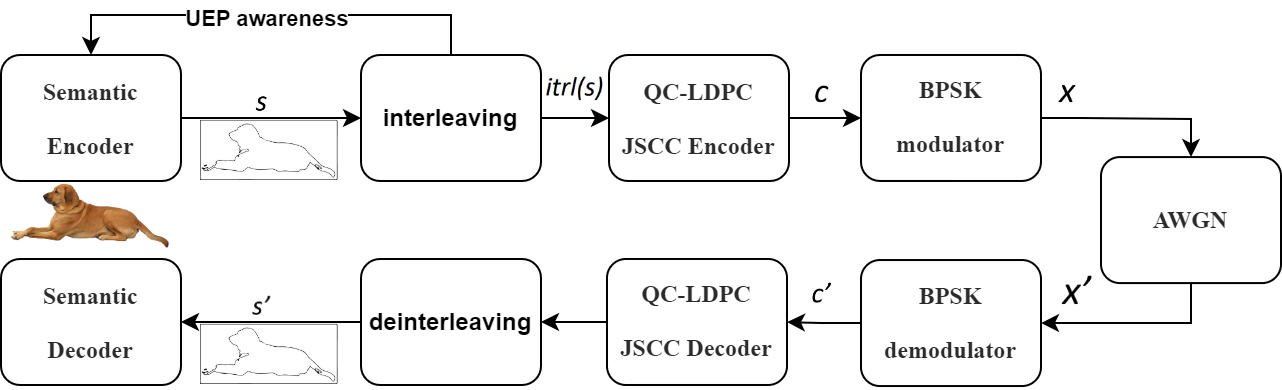}}
\caption{Block diagrams of the proposed semantic-feature image transmission scheme}
\label{fig-JSCC-schemes}
\end{figure*}

The proposed JSCC system is designed to accommodate various types of semantic communications and task-oriented communications through the application of deep learning techniques. A deep learning model recovers the transmitted data with the semantic or task-oriented feature in accordance with side information derived from knowledge-based and task-effectiveness metrics. As depicted in Fig. \ref{fig-JSCC-schemes}, the architecture employs a specialized semantic encoder to transform the raw data into a source sequence based on knowledge-based side information, denoted as \( \mathbf{s} \). This sequence undergoes a random interleaving process or the proposed UEP installation for the semantic encoder to leverage the UEP capability and be aware of the location of the invulnerable codeword segment, resulting in \( \mathbf{itrl(s)} \). Subsequently, the JSCC QC-LDPC encoder with a code rate of \(0.8\) compresses and encodes \( \mathbf{itrl(s)} \) into a new sequence \( \boldsymbol{C} \) to ensure reliable data transmission. The encoded sequence \( \boldsymbol{C} \) is then modulated using BPSK, where bits \(0\) and \(1\) are mapped to \(+1\) and \(-1\), respectively, yielding \( \boldsymbol{X} \). This modulated sequence is transmitted over an Additive White Gaussian Noise (AWGN) channel, resulting in the received signal \( \boldsymbol{X}' = \boldsymbol{X} + N \), where \( N \) represents the AWGN. Finally, a QC-LDPC-based JSCC decoder processes \( \boldsymbol{X}' \), followed by a de-interleaving step, to produce the estimated source sequence \( \mathbf{s'} \) for the semantic decoder.

\subsection{QC-LDPC codes construction and encoding}

The JSCC system outlined in this paper is designed for implementation on hardware such as FPGAs, requiring a shift from floating-point to fixed-point arithmetic. This change inevitably leads to a predictable decrease in computational accuracy. To counteract this performance loss, additional parity bits are incorporated into the JSCC QC-LDPC encoder. This adjustment ensures that the system maintains a reasonable performance level when deployed on hardware. Furthermore, in our previous work \cite{Chou2018}, we provide a UEP LDPC code construction that shows the best irregular node distribution in accordance with the design of the large-degree variable node, which has a better capability for error correction than the small-degree variable node. From a hardware perspective, a variable node unit with a larger degree and stronger error-correcting capabilities has a higher level of computational complexity than a smaller degree. In a tanner graph, the degree of the check/variable node can be represented as its connecting edge. The authors derive the theoretical analysis to demonstrate the decoding capability of UEP LDPC code. This result can be applied to UEP QC-LDPC code design for semantic communication and apply the following QC-LDPC code construction by determining variable node degree to achieve UEP for the JSCC system. 

The construction of QC-LDPC codes is twofold based on Protograph LDPC (P-LDPC) codes and CPMs replacement for the quasi-cyclic characteristic. Our aim is to construct a QC-LDPC matrix based on $(\mathbf{B}_{s1}, \mathbf{B}_{c1})$. The optimized P-LDPC codes, $(\mathbf{B}_{s1}, \mathbf{B}_{c1})$, deriving from \cite{8314100}, are chosen in that such code pair can achieve a decoding threshold of -2.1 dB, 1.5dB lower than the classic $(\mathbf{B}_{R4JA}, \mathbf{B}_{AR4JA})$ \cite{divsalar2009capacity}. Besides, in general, P-LDPC codes \cite{zhan2022design} also offer rapid encoding and decoding structures and achieve the linear minimum Hamming distance, leading to a better performance in the waterfall region and the error-floor region of BER curves. In what is called the lifting method for $(\mathbf{B}_{s1}, \mathbf{B}_{c1})$, the JSCC matrix can be written as 

\begin{equation}\label{H_first_lifting}
\mathbf{H}_{50\times90} = 
\begin{pmatrix}
    \mathbf{H}_{s(20\times40)} & \mathbf{H}_{L(20\times50)}\\ 
    \mathbf{0}_{s(30\times40)} & \mathbf{H}_{c(30\times50)} 
\end{pmatrix}
\end{equation}

where $\mathbf{0}_{30\times40}$ is an all-zero matrix of size $30\times40$. $\mathbf{H}_{L(20\times40)} = [\mathbf{0}_{20\times30}\mathbf{I}_{20\times20}]$.

The next step is to replace traditional LDPC codes with QC-LDPC codes. The parity check matrix of the QC version of the selected P-LDPC codes can be obtained by replacing ``1"s with CPMs of appropriate shift values and ``0"s with all-zero matrices based on $\mathbf{H}_{50\times90}$ using Golomb-Ruler\cite{fossorier2004quasicyclic}, which can ensure the girth of generated matrices are large enough. 

By lifting $\mathbf{H}_{50\times90}$ with a factor of $z = 160$, the matrix of QC-LDPC codes that we obtain can be denoted by

\begin{equation}\label{H_second_lifting}
\mathbf{H}_{QC(50\times90)} = 
\begin{pmatrix}
    \mathbf{H}_{sQC(20\times40)} & \mathbf{H}_{LQC(20\times50)}\\ 
    \mathbf{0}_{sQC(30\times40)} & \mathbf{H}_{cQC(30\times50)} 
\end{pmatrix}
\end{equation}

Referring to Eq.\ref{H_second_lifting}, the parity check matrices for the JSCC encoding are indicated as $\mathbf{{H}_{sQC}}$ and $\mathbf{{H}_{sQC}}$, which sizes are 3200$\times$6400 and 4800$\times$8000, respectively.

The encoding method can be simply achieved by 

\begin{equation}\label{encoding_with_G}
\mathbf{c} = \mathbf{G}^{T}\mathbf{s}
\end{equation}

where, generation matrix $\mathbf{G}$ can be calculated using parity matrix $\mathbf{{H}_{sQC}}$. Since the large size of the proposed $\mathbf{H}$ matrix requires massive computation, the other equivalent approach is leveraging the right side of Eq.\ref{H_second_lifting} to divide and conquer the JSCC encoding scheme. To generate the output of the JSCC encoder, denoted by $c$, two major steps, source compression and parity-bit generation for channel communication should be followed as below.

\textbf{Source Compression}: The first step can be regarded as compressing data, which is given by 

\begin{equation}\label{encoding_with_another_way_1}
\mathbf{b} = \mathbf{{H}_{sQC}}.
\end{equation}

Based on the size of parity check matrices, the source information vector, denoted as $\mathbf{s}$, has a size of 6400 $\times$ 1. As optimized QC-LDPC codes are designed for $p$ = 0.04, this means that the probability of ``1" in the source vector should be 4\%. In other words, $p = Pr(s_{i} = 1)=0.04, \; i=1,2,...,6400$. Given that $\mathbf{H_{sQC}}$ (3200$\times$6400) is the H matrix for the source codes, the compressed output can be calculated using Eq. (\ref{encoding_with_another_way_1}). The size of the output $\mathbf{b}$ is 3200$\times$1. The source compression ratio $R_{s}$ is equal to $6400/3200 = 2$.

\textbf{Channel Parity-bit Generation}: the subsequent manner is based on the property of LDPC parity check matrix in Eq. \ref{encoding_with_another_way_3}:

\begin{equation}\label{encoding_with_another_way_3}
\mathbf{H}_{QC} \: \mathbf{c} = [ \mathbf{{H}_{1(4800\times4800)})} \: \mathbf{{H}_{2(4800\times3200)}}] \: \mathbf{c} = 0.
\end{equation}

As codeword $\mathbf{c}$ of the JSCC encoder can be denoted as

\begin{equation}\label{encoding_with_another_way_2}
\mathbf{c} = [\mathbf{p}\: \mathbf{b}]^T
\end{equation}

where $\mathbf{p}$ represents the parity bit vector of size $1 \times 4800$. Accordingly, the parity bit vector can be calculated using

\begin{equation}\label{encoding_with_another_way_4}
\mathbf{p} = \mathbf{{H}_{1(4800\times4800)}^{-1}} \mathbf{{H}_{2(4800\times3200)}} \mathbf{b}.
\end{equation}

The second half of the proposed JSCC encoder, which behaves like a channel encoder and is fed with a 3200-bit long$\mathbf{b}$, outputs 8000 bits. Therefore, the channel code rate $R_{c}$ is $3200/8000 = 0.4$. So far, the overall code rate will be $ R_{overall} = R_{s} \times R_{c} = 0.8$.

It is noted that all data involved in the JSCC encoding scheme are represented as binaries, corresponding operations depicted in the aforementioned equations are merely bitwise operators (ANDs and XORs), which are hardware-friendly to implement.

\subsection{JSCC Layered decoding algorithm}

\begin{figure*}[htbp]
\centerline{\includegraphics[width=1\textwidth]{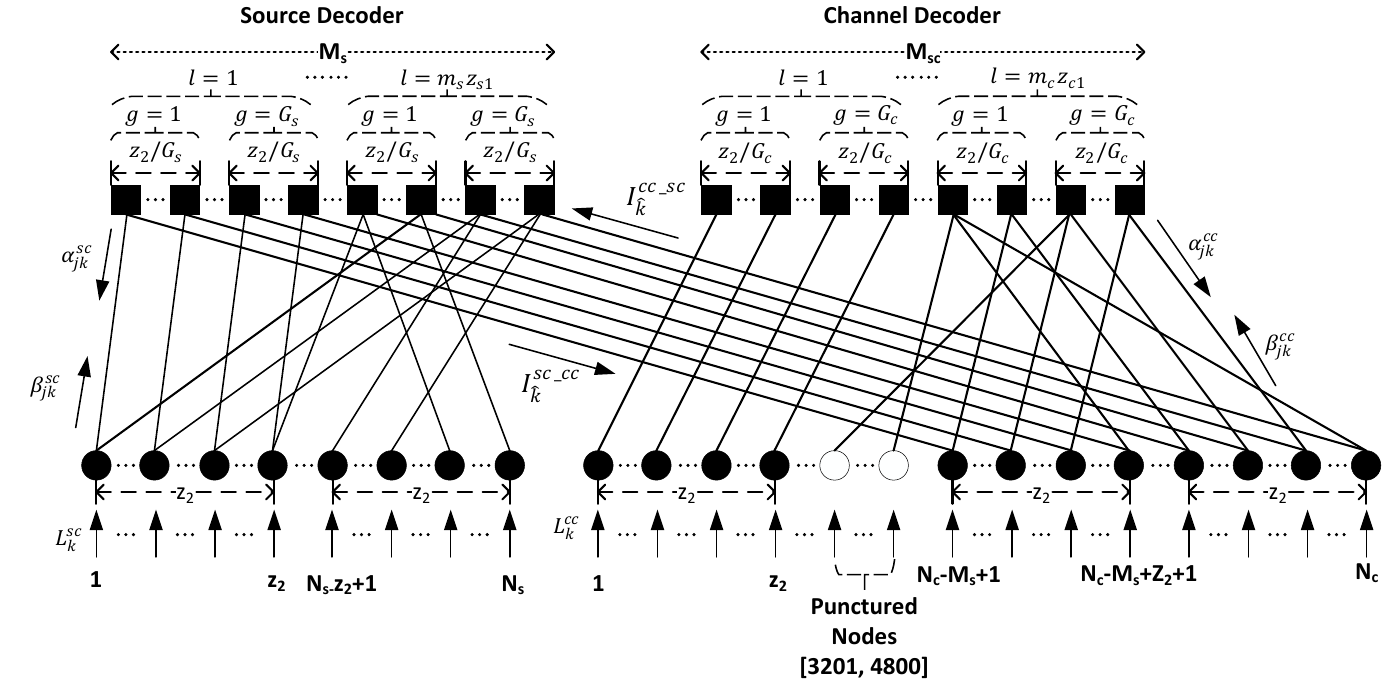}}
\caption{The Bipartite Graph of JSCC Decoding Scheme using QC-LDPC codes}
\label{fig-JSCC-bipartite}
\end{figure*}
The JSCC decoder using QC-LDPC codes can be visualized through a Tanner graph. This graph is essentially divided into two interconnected subgraphs: one representing the source and the other representing the channel, depicted in Fig. \ref{fig-JSCC-bipartite}. As the major component of the proposed system, the QC-LDPC code-based JSCC decoder can be achieved by applying a QC-LDPC layered sum-product decoding algorithm\cite{hu2001efficient}\cite{gunnam2007vlsi}, although some new calculations, such as message exchanges between the source side and the channel side, need to be included. The partial parallelism of the layered decoding algorithm, which enables full parallel operations amongst all sub-matrices within one check node or ``layer", simplifies the hardware implementation of the source and channel decoders. The following parameters and values should be defined and assumed before explaining the JSCC decoding process.
\begin{itemize}
    \item An AWGN channel with zero mean and variance $\delta^{2}$ is assumed. The channel value from this AWGN channel is set as the initial values of Variable-Node-to-Check-Node (V2C) messages.
    \item $M_{s}$ and $M_{c}$ represent the number of check nodes (CNs) in the source and the channel, respectively.
    \item $N_{s}$ and $N_{c}$ represent the number of variable nodes (VNs) in the source and the channel, respectively.
\end{itemize}

The key decoding procedures for the source and channel are as follows.

\begin{itemize}
\item Variable node processor (VNP) in source decoding: the variable node to check node (V2C) messages,$\beta _{jk}^{sc}$, is computed according to Eq. (\ref{Chpt3:sc_vnp_equation}). To be more specific, $L_{sc}^{k}$ is denoted as the source log-likelihood ratio (LLR) of the $k$-th VN in the source decoder. $M(k)\setminus j$ represents the set of CNs connected to the $k$-th VN, excluding the $j$-th CN itself. The other denotations regarding all other equations can be found in Fig.\ref{fig-JSCC-bipartite}. It should be noted that circles and squares, respectively, represent variable nodes and check nodes in Fig. \ref{fig-JSCC-bipartite}. The hollow circles represent punctured variable nodes.
\end{itemize}

\begin{equation}\label{Chpt3:sc_vnp_equation}
\beta _{jk}^{sc} = L^{sc}_{k}+\sum_{{j}'\in M(k)\setminus j}^{}\alpha _{jk}^{sc} \;\;\;\;\;\;\;\;\;\forall k\in N(j)  
\end{equation}

\begin{itemize}
\item Check node processor (CNP) in source decoding: Two operations are performed in two steps.
\end{itemize}
\begin{enumerate}
    \item  Update the check node to variable node (C2V) messages $\alpha_{jk}^{sc}$ in Equation \ref{sc_cnp_equation}. 
    \item  Updating message from the source decoder to the channel decoder, $I_{\hat{k}}^{sc\_cc}$ in Equation \ref{sc_cc_equation}.
\end{enumerate}

\begin{equation}\label{sc_cnp_equation}
\begin{split}
tanh(\alpha_{jk}^{sc}/2) = tanh(I^{cc\_sc}_{\hat{k}}/2) \times \prod_{{k}'\in N(j)\setminus k}^{}tanh(\beta_{jk}^{sc}/2) \\
\forall k\in N(j)  
\end{split}
\end{equation}\

\begin{equation}\label{sc_cc_equation}
tanh(I_{\hat{k}}^{sc\_cc}/2) = \prod_{{k}'\in N(j)}^{}tanh(\beta_{j{k}'}^{sc}/2) 
\end{equation}

Calculating \textit{Posteriori} LLR $l_{k}^{sc}$ needs to be done after CNP using Eq. (\ref{sc_llr_equation}).

\begin{equation}\label{sc_llr_equation}
l_{k}^{sc}=L_{k}^{sc}+\sum_{{j}'\in M(k)}^{}\alpha_{{j}'k}^{sc}.     
\end{equation}

For channel decoding, the procedure for updating LLR messages is very similar to source decoding, as depicted on the right side of Fig. \ref{fig-JSCC-bipartite}.

\begin{itemize}
\item VNP in channel decoding: owing to the crossed message-passing mechanism between the source decoder and the channel decoder, VNP in channel decoding must be computed separately. $\widetilde{N}^{cc\_sc}$ indicates the set of VNs connected to the corresponding CNs in the source decoder.
\end{itemize}

\begin{equation}
\beta_{jk}^{cc} = L_k^{cc} + \sum_{{j}'\in M(k)\setminus j}^{}\alpha_{{j}'k}^{cc}  \;\;\;\;\;\;\;\;\;\forall k\in N(j)\cap \widetilde{N}^{cc\_sc} \label{vnp_cc_1}
\end{equation}

In Eq. (\ref{vnp_cc_1}), $\widetilde{N}^{cc\_sc}$is the complement of $N^{cc\_sc}$.

\begin{equation}
\beta_{jk}^{cc} = I_{\hat{k}}^{sc\_cc} + L_k^{cc} + \sum_{{j}'\in M(k)\setminus j}^{}\alpha_{{j}'k}^{cc}  \;\;\;\;\;\;\;\forall k\in N(j)\cap N^{cc\_sc} \label{vnp_cc_2}
\end{equation}

\begin{itemize}
\item CNP in the channel decoding:
\end{itemize}
\begin{equation}
tanh(\alpha_{jk}^{cc}/2) = \prod_{{k}' \in N(j)\setminus k}^{}tanh(\beta_{jk}^{cc}/2)\;\;\;\;\;\;\;\;\;\forall k\in N(j) \label{cc_cnp_eq}
\end{equation}

\begin{itemize}
\item Message from the channel decoder to the source decoder:
\end{itemize}

\begin{equation}
I_{\hat{k}}^{cc\_sc} = L_{\hat{k}}^{cc} + \sum_{{j}' \in M(\hat{k})}^{}\alpha_{{j}'\hat{k}}^{cc}. \label{I_cc_sc_eq}
\end{equation}

\begin{itemize}
\item Posteriori LLR update without VNs connected with CNs in the source decoder: 
\end{itemize}

\begin{equation}
l_k^{cc}=L_K^{cc} + \sum_{{j}'\in M(k)}^{} \alpha_{{j}'k}^{cc} \;\;\;\;\;\;\;\;\;\forall k\in N(j)\cap \widetilde{N}^{cc\_sc} \label{updateLLRa}
\end{equation}

\begin{itemize}
\item A Posteriori LLR update with only VNs connected with CNs in the source decoder:
\end{itemize}

\begin{equation}
\begin{split}
l_{\hat{k}}^{cc}=I_{\hat{k}}^{cc}+L_{\hat{k}}^{sc\_cc}+\sum_{{j}' \in M(\hat{k})}^{}{\alpha_{{j}'\hat{k}}^{cc}}\;\;\;\;
\forall k\in N(j)\cap N^{cc\_sc}   \label{updateLLRb}
\end{split}
\end{equation}

The last step is the hard decision and the stopping criterion for the decoding iteration. Let us consider the decoded source-side sequence and channel-side sequence with $\hat{s} = \{\hat{s}_1, \hat{s}_2, ..., \hat{s}_k\}$ and $\hat{c} = \{\hat{c}_1, \hat{c}_2, ..., \hat{c}_k\}$, respectively. They are used to obtain an estimated value of the received codeword sent on the sender's side, according to the following rule:

\begin{itemize}
\item $\hat{c}_k$ = 0 if $l_k^{cc} \geq $ 0, otherwise $\hat{c}_k$ = 1 $\forall k$
\item $\hat{s}_k$ = 0 if $l_k^{sc} \geq $ 0, otherwise $\hat{s}_k$ = 1 $\forall k$
\end{itemize}

Stopping the decoding process is subject to if the decoder reaches the maximum number of decoding iterations, which is preset before decoding.

\begin{figure*}[htbp]
\centerline{\includegraphics[width=1\textwidth]{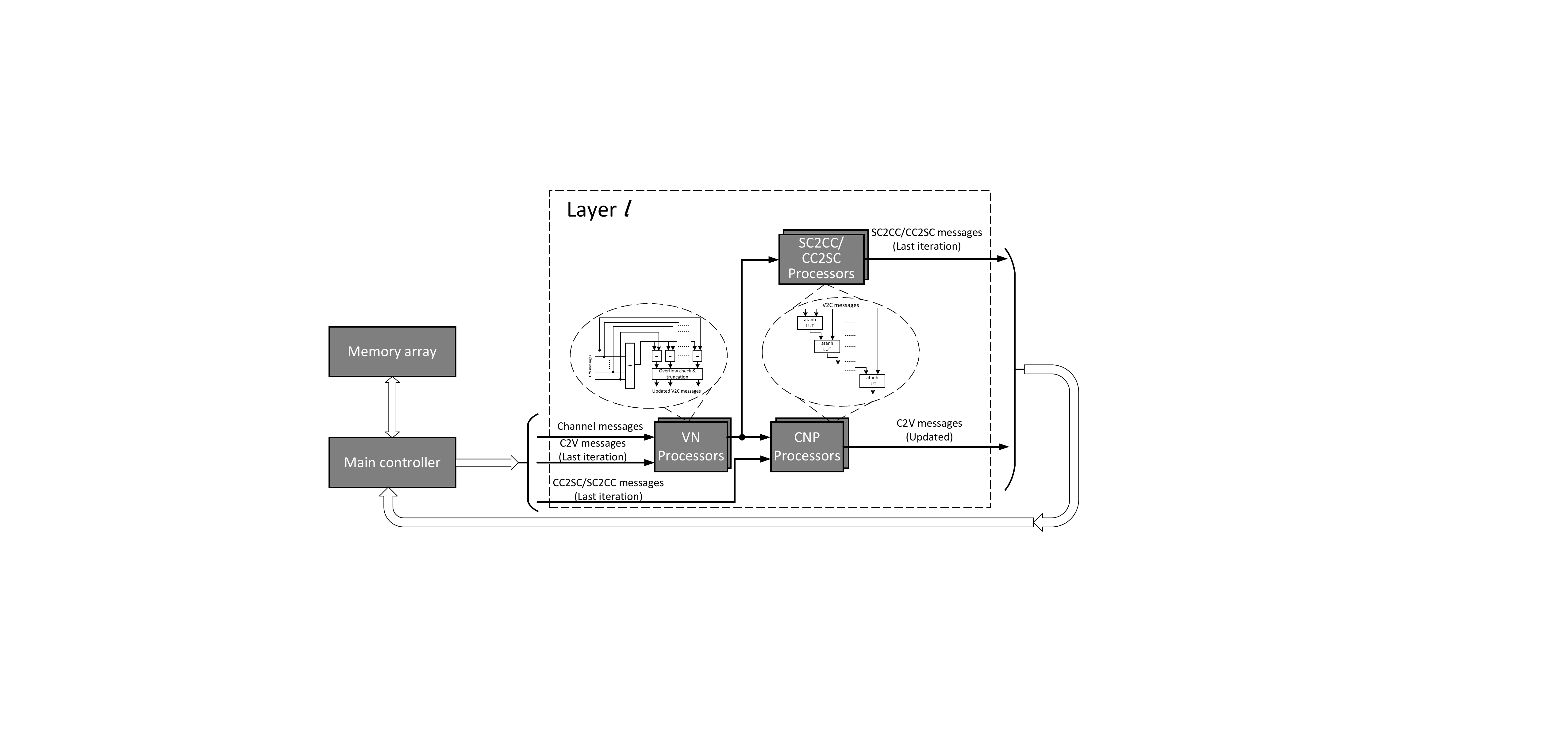}}
\caption{A brief design architecture of the QC-LDPC decoder implementation for either the Source or Channel side.}
\label{fig-JSCC-decoder}
\end{figure*}

It should be emphasized that in our case, the number of layers is: $m_{s}z_{s1}$ = 20. $G_{s}$ and $G_{c}$, representing the number of decoding groups per layer in the source and channel decoder, respectively, and we choose one group ($G_{s}$ = 1 and $G_{c}$ = 1) for both sides.

The proposed architecture shown in Fig. \ref{fig-JSCC-decoder} is different from a normal layered LDPC decoding architecture, as $I^{cc\_sc}_{\hat{k}}$ and $I_{\hat{k}}^{sc\_cc}$, used for exchanging messages between the source and channel decoders, in Eq. \ref{I_cc_sc_eq} and Eq. \ref{sc_cc_equation} are involved exclusively in a JSCC system. Therefore, $I^{cc\_sc}_{\hat{k}}$ and $I_{\hat{k}}^{sc\_cc}$ should be calculated as well in the CC2SC/SC2CC Processors. To balance the complexity of this system, a modified quantized sum-product algorithm based on \cite{6419075} was adopted to simplify the hyperbolic tangent functions in Eq. \ref{sc_cnp_equation} and Eq. \ref{cc_cnp_eq}. Specifically, a look-up table (LUT) architecture is implemented for a two-input fixed-point $tanh$ calculation.

\subsection{Interleaver and De-interleaver for UEP installation}
\begin{figure}[htbp]
\centerline{\includegraphics[width=0.5\textwidth]{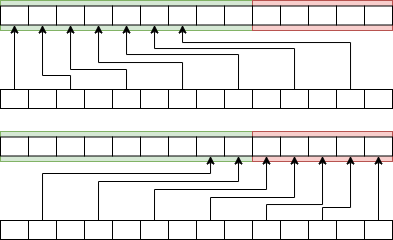}}
\caption{The proposed regular interleaver in \cite{Chou2017} for UEP intallation}
\label{fig-JSCC-inter}
\end{figure}
The interleaving technique is used to spread burst errors and average the distribution of ``0"s and ``1"s in the source vector. Especially for semantic feature images, the importance is likely to be centered and continuous. Interleaving the information in these images can improve the error correction capability of LDPC codes. To recover the order of each binary on the decoding side, the de-interleaver is applied. In \cite{Chou2017}, the proposed regular interleaver and de-interleaver are shown in Fig.\ref{fig-JSCC-inter} with the assumption of semantic importance having a higher occurrence probability on the even position using the signal processing technique or semantic model. This technique can be designed for the allocation of semantic importance to invulnerable or stronger error-correction capability as illustrated in the red codeword segment in contrast to the vulnerable codeword in the green segment. Hence, the semantic encoder/decoder possesses information on the distribution of UEP in the transmitted coded sequence. Consequently, it may achieve enhanced semantic compression while semantic importance is protected by an invulnerable codeword segment.

\begin{table}[htbp]
\centering
\caption{Characters of the synthesized JSCC decoder only}
\label{table:utilization_0}
\begin{tabular}{cccc}
\toprule  
 & Fix-point Implementation on FPGA \\
\midrule  
Quantization bit width       & 6 \\
H Matrix size (Channel) & 8000x4800 \\
H Matrix size (Source) & 6400x3200 \\
Sub-matrix size (both) & 160 \\
BRAM 18K     & 286  (1.4\%) \\
FF           & 214K  (33\%) \\
LUT          & 643K  (33\%) \\
Latency per iteration  & 31ms \\
Clock Rate(MHz)    & 100    \\
\bottomrule 
\end{tabular}
\end{table}

\section{Performance Evaluation}

\subsection{Experimental Platform}

\begin{figure*}[htbp]
\centerline{\includegraphics[width=\textwidth]{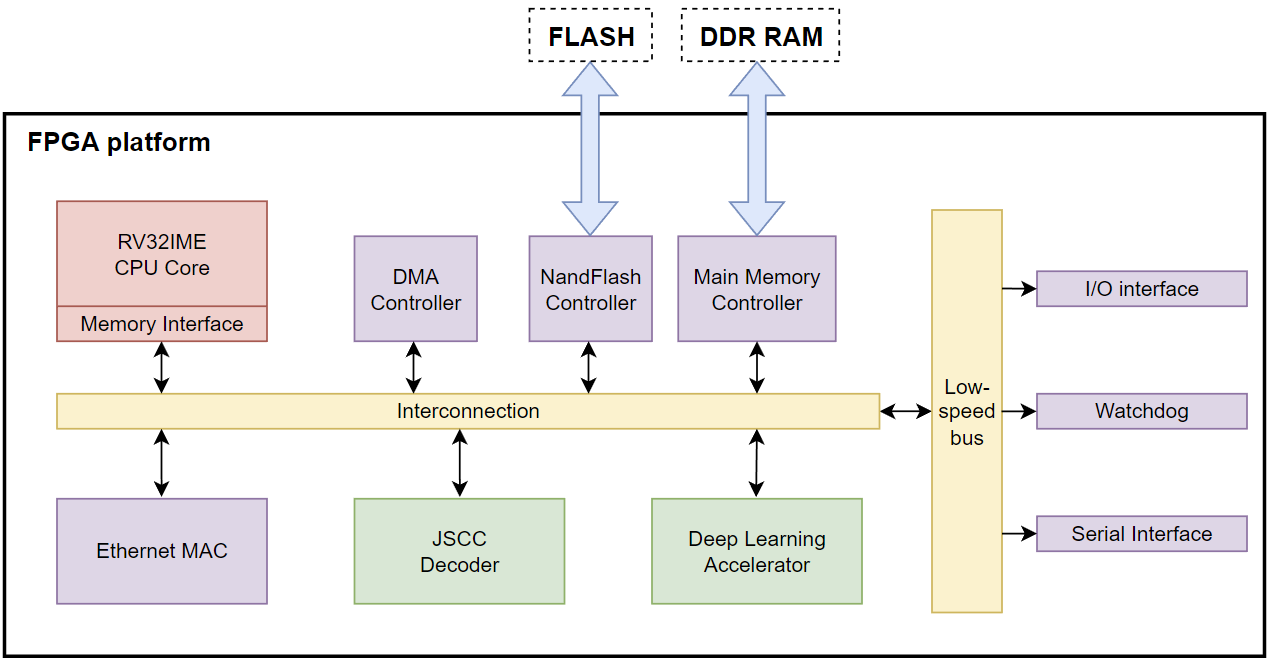}}
\caption{The Experimental Platform on the receiver's side}
\label{JSCC_system_arch}
\end{figure*}

The experimental setup utilizes a Virtex Ultrascale + FPGA VCU118 evaluation kit as the receiving platform, shown in Fig. \ref{JSCC_system_arch}. The system's processor is built around an open-source 32-bit RISC-V CPU core, which is connected to a flexible interconnection bus. Essential modules such as DMA controllers, NAND flash controllers, and main memory controllers for DDR RAMs are integrated, along with an Ethernet IP. The core components of the system include a specialized JSCC decoder based on QC-LDPC codes and a deep learning accelerator for semantic and task-oriented processes. This accelerator is designed to handle matrix multiplication, convolution, and activation functions, which are fundamental to neural network computations based on YOLO\cite{JIANG20221066} and auto-encoder. To simulate transmission modules (in the first row depicted in Fig. \ref{fig-JSCC-schemes}) and model an AWGN channel, a separate computer is employed. Consequently, the combined computer-FPGA setup functions as two distinct JSCC systems. The first system encodes semantic feature images and transmits them as BPSK-modulated data through an AWGN channel. The second system receives these image data from the channel and reconstructs the original image information using demodulation and JSCC decoding techniques.

\subsection{Implementation results}

A 6-bit quantization scheme denoted as "Proposed-Q6" in Fig. \ref{table:imp_character_with_others}, is adopted in the proposed JSCC system. The synthesized results for the main component, the QC-LDPC code-based JSCC decoder, are shown in Table \ref{table:utilization_0}. It should be noted that the logic (FFs and LUTs) to implement the decoder occupies around 33\% of the entire FPGA and around 1.4\% of block memories. For each decoding iteration, 31 milliseconds are spent by the FPGA driven by a 100 MHz clock. In Table \ref{table:imp_character_with_others}, the feature comparison reveals that our FPGA platform for JSCC semantic communication encompasses both floating point analysis and fixed point (6-bit) quantization. Additionally, the inclusion of supplemental parity bits enhances the BER performance when employing a code rate of 0.8 to compensate for the quantization loss, in comparison to the other platforms utilizing a code rate of 1.

\begin{table}[htbp]
\caption{Feature Comparison with others}
\label{table:imp_character_with_others}
\begin{tabular}{llll}
\hline
                 & Proposed-FP32    & Proposed-Q6  & \begin{tabular}[c]{@{}l@{}}others\cite{e23111392,9115063}\\ \cite{8314100,7452543,dong2022joint}\end{tabular} \\    
\hline
Data type        & FP 32bits        & \textbf{Fixed 6 bits} & FP 32bits     \\
ECC type         & \textbf{QC-LDPC} & \textbf{QC-LDPC}            & LDPC          \\
Implemented?     & No               & \textbf{Yes}                & No            \\
Code rate        & 0.8              & 0.8                         & \textbf{1.0}  \\ 
\hline
\end{tabular}
\end{table}


\begin{figure}[htbp]
\centerline{\includegraphics[width=0.5\textwidth]{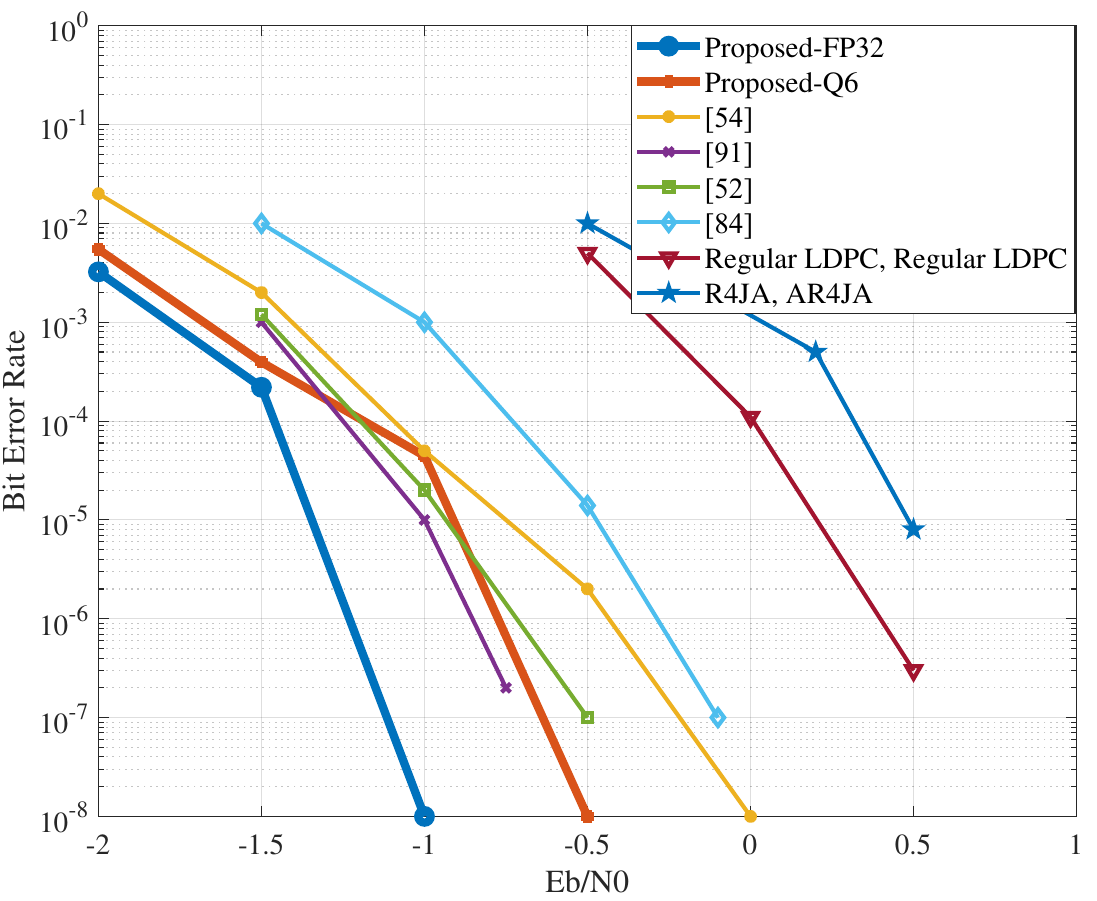}}
\caption{BER performance comparison. $p=0.04$}
\label{fig-JSCC-BER-perf}
\end{figure}

Owing to the distinct architecture of the proposed scheme and the dimensions of the LDPC matrix, a direct performance comparison with extant literature is not feasible. As a result, a simulated system employing 32-bit floating-point precision, labeled as 'Proposed-FP32' in Fig. \ref{table:imp_character_with_others}, serves as a benchmark against the 6-bit hardware implementation and six additional relevant studies. As illustrated in Fig. \ref{fig-JSCC-BER-perf}, the Bit Error Rates (BER) for the proposed codes, both under floating-point simulation and 6-bit hardware implementation, were scrutinized in juxtaposition with other works that solely investigated JSCC systems in a simulated setting, as indicated in the final column of Fig. \ref{table:imp_character_with_others}. The BER is plotted on the y-axis, while the $E_{b}/N_{0}$, or the energy per bit to noise power spectral density ratio, is plotted on the x-axis. The empirical findings from the 6-bit quantized version align well with the simulated outcomes, denoted as 'Proposed FP32.' Notably, the trajectory of these curves, as $E_{b}/N_{0}$ increases, is congruent with existing scholarly contributions. Despite the simplifications introduced by data quantization in the proposed architecture, a modest decline in the BER of 'Proposed-Q6' is observed. However, its BER performance remains superior to those reported by Z.Xu et al. \cite{e23111392}, Q.Chen et al. \cite{7452543}, Double regular LDPC, and R4JA in \cite{dong2022joint}, all of which employ non-QC-LDPC codes with 32-bit floating-point simulations. This superior performance can primarily be attributed to the QC-LDPC code generation, reasonable code rate modifications, and efficacious hardware deployment.



\begin{figure}[htbp]
\centerline{\includegraphics[width=0.5\textwidth]{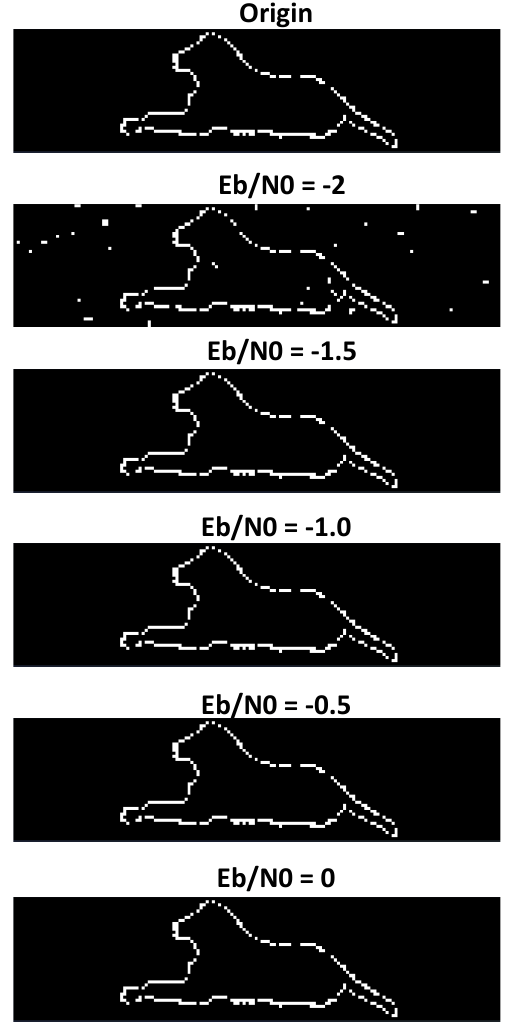}}
\caption{Original and Received Semantic Feature Images in a case when $p=0.0398$}
\label{fig-JSCC-Sig-Image-Output}
\end{figure}

As the QC-LDPC codes in this design are optimized on the condition that $p \leq 0.04$, a semantic feature, such as the original one depicted in the upper image of Fig. \ref{fig-JSCC-Sig-Image-Output}, can be well contained in a 160$\times$40 image, in which each black and white pixel can be represented by 1 bit in the source vector. This compact system is tested under $E_{b}/N_{0}$ ranging from $-2$ to $0$ with 0.5 as its step. Fig. \ref{fig-JSCC-Sig-Image-Output} shows one occasion under various $E_{b}/N_{0}$. Considering the low BER results in Fig. \ref{fig-JSCC-BER-perf}, this simple semantic feature sender and receiver can communicate correctly with each other nearly all the time. The code rate for this entire system is $R = R_{s} \times R_{c} = 2 \times 0.4 = 0.8$. It should be noted that in such low $E_{b}/N_{0}$, many communication standards, such as IEEE 802.11n (WiFi) and IEEE 802.16e (WiMAX), need to set a low code rate to 0.5 or even lower when complex or high-volume noise is detected.

\section{Conclusion and Future Directions}
\label{sec:conclusion}
By facilitating the exchange of highly informational, up-to-date, and efficient data. SC has the potential to enhance the effective use of resources, improve information accuracy and efficacy in task completion, and serve as a model and technical foundation for future generations of communication systems. The utilization of task-oriented communication has been widely regarded as a novel approach in the development of communication methods for multi-agent systems. In this paper, a novel JSCC system based on QC-LDPC codes is proposed as a promising candidate for semantic task-oriented communication systems. As the proposed irregular QC-LDPCcode construction with the nature of UEP capability, the semantic importance can be dynamically assigned to the variable node with stronger error-correction capability by the proposed interleaver. After semantic source coding passes the information to the simple structure of the QC-LDPC codes, the JSCC system is implemented on the hardware device. Significantly, the operations of the JSCC decoder are layered and are then executed in parallel both on the source and channel sides. The fixed-point system also maintains fair BER performance compared to the simulated one. Moreover, the design with its optimized QC-LDPC codes is further investigated by compressing image data and protecting encoded data via an AWGN channel. This application is the semantic feature of image transmission and reception. In many cases, in practice, sources are transmitted uncoded, while state-of-the-art block channel encoders are used to protect the transmitted data against channel errors. If the JSCC scheme is used in such cases, the throughput can be improved by compressing the source data ($R_{s} = 2$) and then the channel coding starts. If the input image and channel codes are fixed, the throughput of our proposed design is doubled, compared with the non-compression system. Even when competing with another JSCC system, a better error correction capability can still outperform others. We conclude some potential future research directions as follows. 
\begin{enumerate}
\item The design concepts entail the utilization of collaborative techniques with an edge AI server throughout the process of semantic transmission in order to enhance adaptability. The objective of this collaboration is to provide guidelines for configuring the JSCC decoder in the context of feature and federated learning.
\item Due to the nature of the applicable implementation in \cite{pham2023binarizing}, the JSCC framework is used to interface with binarized neural networks to optimize resource utilization in edge devices with limited computational capabilities. This approach is specifically applied in the context of task-oriented communication and goal-oriented quantization while considering the application of the RISC-V low-power revolution.
\item Although there is a tendency for DJSCC to supplant the role of JSCC, the future possibility of collaboration with machine learning remains a focal point not only in the study of theoretical analysis but also in practical implementation. The importance of including semantic and task-oriented design aspects in the development of UEP JSCC prototype cannot be overstated, particularly in the context of edge AI for future-generation 6G communications.
\item  In the cybersecurity aspect, the analysis of the secrecy rate transition between edge servers and devices utilizing JSCC can be explored within the framework of differential privacy\cite{liu2021privacy}, specifically in the presence of eavesdropping or the other assault scenario.
\end{enumerate}
The incorporation of semantics and task-oriented communication is expected to assume a significant role in forthcoming intelligent systems. The purpose of this article is to offer an introductory overview and a cohesive perspective for the prototype of next-generation 6G communication systems to inspire more potential research activities.

\bibliographystyle{IEEEtran}
\balance
\bibliography{main}


\end{document}